\documentclass[conference]{IEEEtran}
\IEEEoverridecommandlockouts
\usepackage{cite}
\usepackage{amsmath,amssymb,amsfonts}
\usepackage{algorithmic}
\usepackage{graphicx}
\usepackage{textcomp}
\usepackage{xcolor}
\usepackage{algorithm}
\usepackage{subcaption} 
\usepackage{amsmath, bm}
\usepackage{subcaption}
\def\BibTeX{{\rm B\kern-.05em{\sc i\kern-.025em b}\kern-.08em
    T\kern-.1667em\lower.7ex\hbox{E}\kern-.125emX}}
\begin{document}
\renewcommand{\refname}{REFERENCES}

\title{ Flag-Preamble-Based Delay-Doppler Channel Estimation for Next-Evolution Waveforms\\
\thanks{This work was supported by the Shenzhen Science and Technology
Major Project under Grant KJZD20240903102000001 and the Science and
Technology Planning Project of Key Laboratory of Advanced IntelliSense
Technology, Guangdong Science and Technology Department under Grant
2023B1212060024. Email: \{liuy2659@mail2, tangyq8@mail\}.sysu.edu.cn.}
}

\author{\IEEEauthorblockN{Yuan Liu$^1$, Yilin Qi$^1$, Caiqin Li$^{1,2}$, Haoran Yin$^1$, Yanqun Tang$^{1,3}$ and Yao Ge$^{4}$}
	\IEEEauthorblockA{
		${}^1$\textit{School of Electronics and Communication Engineering, Sun Yat-sen University, China}
        \\
        ${}^2$\textit{School of Computer and Communication Engineering, Northeastern University at Qinhuangdao, China}
        \\
        ${}^3$\textit{Guangdong Provincial Key Laboratory of Sea-Air-Space Communication, China}
        \\
        ${}^4$\textit{AUMOVIO-NTU Corporate Lab, Nanyang Technological University, Singapore}
        \\
}}
\maketitle

\begin{abstract}
Accurate delay-Doppler channel estimation is critical for next-evolution waveforms (NEWs) to enable reliable signal detection. This paper proposes a robust channel estimation algorithm that integrates Flag sequences optimized via an adaptive accelerated parallel majorization-minimization (AP-MM) algorithm with a proposed channel estimation algorithm. 
To enable efficient, low-complexity parameter extraction and further overcome the robustness issues of conventional greedy estimation, we introduce two key enhancements, i.e., a candidate selection strategy to mitigate spurious sidelobe peaks, and a global least squares (LS) refinement stage to eliminate error propagation caused by sidelobe masking effects. Numerical results demonstrate that the proposed scheme significantly outperforms traditional existing algorithms, achieving the desired estimation accuracy.
\end{abstract}


\section{INTRODUCTION}
Next-generation wireless systems aim to support ubiquitous connectivity in highly dynamic environments characterized by significant delay and Doppler spreads, where conventional orthogonal frequency division multiplexing (OFDM) suffers from severe inter-carrier interference (ICI). To mitigate these impairments, several next-evolution waveforms (NEWs), e.g., orthogonal time-frequency space (OTFS), orthogonal delay-Doppler division multiplexing (ODDM), affine frequency division multiplexing (AFDM), and frequency-modulated OFDM (FM-OFDM), have emerged as promising candidates due to their resilience against Doppler spread caused by high mobility\cite{b1,b2}. 

One of the biggest challenges in communications over high-mobility scenarios is how to acquire accurate channel state information (CSI) of the time-varying channels. The authors in \cite{b3,b4,b5} proposed to embed a pilot among data symbols to estimate the delay-Doppler channel, which inevitably introduce pilot-data interference. The authors in \cite{b6} and \cite{b7} proposed to apply advanced compressed sensing (CS) algorithms to leverage channel sparsity, which, however, necessitate large sensing dictionaries and intensive matrix operations. In contrast, the authors in \cite{b8,b9,b10} investigated matched filter (MF) approaches to determine the delay-Doppler taps of the channel, where the accuracy can be fundamentally characterized by the ambiguity function (AF) of the transmitted signal. However, exhaustive searches over fine-grained grids result in $O(N^3)$ or $O(N^2 \log N)$ complexity, where $N$ denotes the sequence length, imposing prohibitive latency for wideband systems. To tackle this issue, the Flag sequence and Flag algorithm, proposed in \cite{b11}, reconcile the conflict between accuracy and complexity by constructing Heisenberg-Weil sequences (HWS) with a unique peak-curtain AF. It decouples the expensive two-dimensional search into two rapid one-dimensional steps, reducing the complexity to a near-linear $O(N \log N)$, particularly suitable for practical channel estimation in high-mobility communications. 

This paper delves into the practical channel estimation algorithm in the delay-Doppler domain based on the Flag preamble. In consideration of the computational constraints in high-mobility communications, Flag sequences are adopted as the preamble structure to ensure rapid parameter extraction. We further propose a practical estimation algorithm with a candidate selection strategy and a global least squares (LS) refinement strategy to obtain precise channel estimation. To this end, we can significantly reduce the probability of false alarms caused by spurious peaks and accurately recover complex gains to mitigate error propagation. 
In addition, simulation results demonstrate that our proposed algorithm can provide accurate CSI for AFDM and can be extended to other promising NEWs.

The remainder of this paper is organized as follows. Section II introduces the preliminaries and system model. Section III introduces the traditional Flag algorithm and the proposed Flag algorithm for channel estimation, as well as complexity analysis. Section IV provides the simulation results, followed by the conclusion in Section V.

\textit{Notation:} $(\cdot)^T$ and $(\cdot)^H$ stand for the transpose and conjugate transpose vectorization of a matrix, respectively. $\text{diag}(\boldsymbol{\rho})$ is a diagonal matrix constructed with elements of $\boldsymbol{\rho}$ as its principal diagonal. $\|\cdot\|$ represents the $l_2$-norm of a vector. $(\cdot)^*$, $|\cdot|$, and $\arg(\cdot)$ denote the conjugate, absolute value, and phase of a complex number, respectively. The imaginary unit is denoted by $j = \sqrt{-1}$. $\mathbf{I}_M$ represents the $M \times M$ identity matrix. $\odot$ denotes the element-wise product.

\section{PRELIMINARIES AND SYSTEM MODEL}
In this section, we first introduce the basic concept of the Flag sequence. Then, we briefly present an AFDM-based system model considered in our work as an example.

\subsection{Flag Sequence}
The Flag sequence, composed of a Curtain sequence $\textbf{{\textit{c}}}$ and a Peak sequence $\textbf{{\textit{p}}}$, can be expressed as
\begin{equation}
\textbf{\textit{f}}=(\textbf{\textit{c}}+\textbf{\textit{p}})/{\sqrt2}.
\label{eqyhr02.3}
\end{equation}

\begin{figure}[t]
    \centering
        \includegraphics[width=0.4\textwidth]{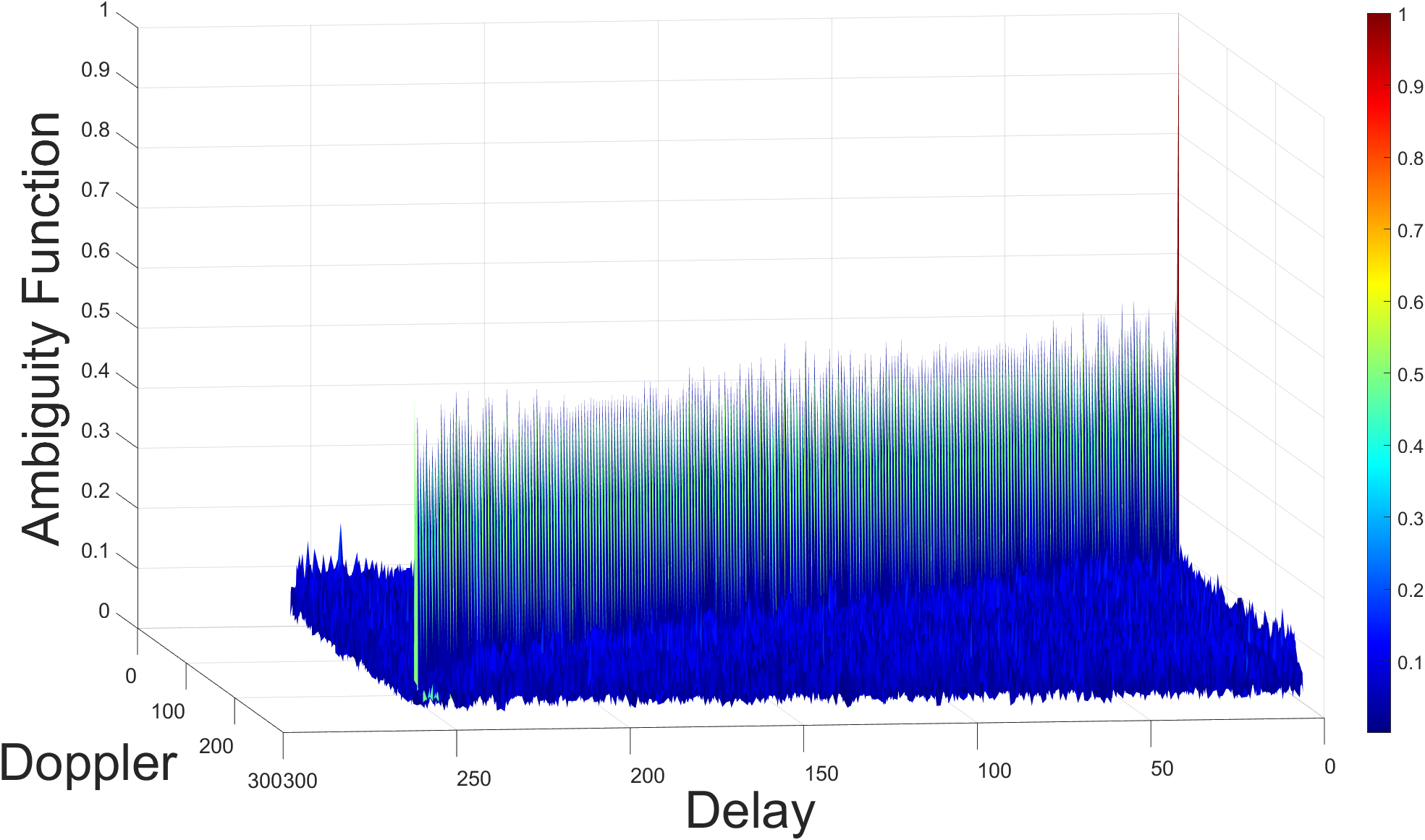}
        \caption{Ambiguity function of traditional Flag sequence. }
        \label{fig:AF_FINAL}
        \vspace {-0.5 cm}
\end{figure}

The authors in \cite{b11} employed a Heisenberg sequence as the Curtain sequence, to focus energy along a specific line $L$ passing through the origin in the delay-Doppler plane. Meanwhile, a Weil sequence is adopted as the Peak sequence to ensure that the energy is concentrated at a single point in the line. The Fig.\ref{fig:AF_FINAL} is the AF of the HWS-based Flag sequence, which has a length of 257.

However, the traditional Flag sequence based on HWS in \cite{b11} is constrained by prime-length requirements and focuses on periodic properties. 
To overcome HWS limitations, the authors in\cite{b12,b13} proposed a novel Flag sequence design. The Curtain sequence is formulated as a discrete chirp sequence

\begin{equation}
\textit{\textbf{c}}_{\xi,q}[n] = \frac{1}{\sqrt{N}} \exp\left( \frac{j n (\xi n + q) \pi}{N} \right),
\label{eqyhr02.3}
\end{equation}
where $\xi$ and $q\in[1-N,2-N,...,N-1]$ are the chirp rate and phase shift index, respectively. The algebraic condition $[\xi N - q]_2 = 0$ ensures an ideal curtain shape for arbitrary lengths $N$. 

Furthermore, due to the sidelobes of the Peak sequence, the authors in \cite{b13} introduce an optimization algorithm to tackle this issue. Unlike the fixed structure of HWS, the Peak sequence and receiver filters are jointly optimized by minimizing the weighted integrated masked sidelobe level (WImSL) objective function via a novel adaptive accelerated parallel majorization-minimization (AP-MM) algorithm. For the specific algorithm process, one can refer to \cite{b12,b13}.

\subsection{System Model}
\begin{figure}[t]
    \centering
        \includegraphics[width=0.5\textwidth]{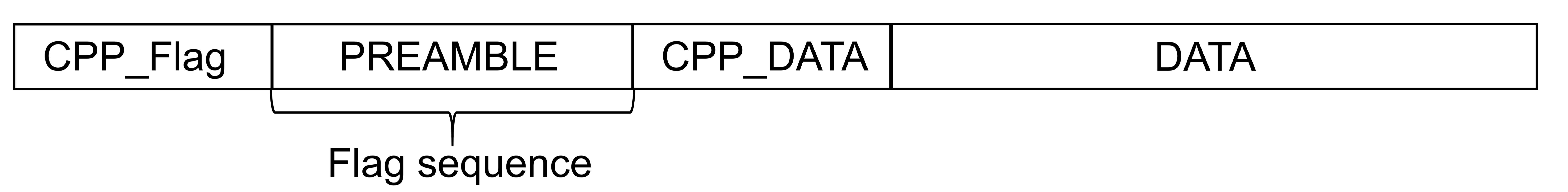}
        \caption{Frame structure. }
        \label{fig:fig_frame}
        \vspace {-0.5 cm}
\end{figure}

This subsection formulates an integrated communication system model, including the Flag preamble sequence and AFDM transceiver model. As the Flag preamble sequence is highly universal, this system model is also applicable to other multi-NEWs, like FM-OFDM\cite{b14}, OCDM\cite{b15}, OTFS\cite{b16,b17}. In this communication system model, we take the Flag sequence as the preamble to estimate channel parameters, followed by the data transmission. The frame structure is illustrated in Fig.\ref{fig:fig_frame}.

Let $\mathbf{x}\in \mathbb{A}^{N\times1}$ denote the vectorized transmitted information symbols and $\mathbb{A}$ represents the constellation set. After $N$-point inverse discrete affine Fourier transform (IDAFT), the information symbols are converted from the discrete affine Fourier transform (DAFT) domain to the time domain, represented as\cite{b18,b19,b20}
\begin{equation}
s[n] = \frac{1}{\sqrt{N}} \sum_{m=0}^{N-1} x[m] e^{j2\pi \left( c_1 n^2 + \frac{1}{N} m n + c_2 m^2 \right)},
\label{eqyhr02.4}
\end{equation}
where $n=0,1,...,N-1$, $c_1$ and $c_2$ are the chirp parameters to be optimized for full diversity\cite{b6}. In matrix form, equation (\ref{eqyhr02.4}) can be written as
\begin{equation}
\mathbf{s} = \mathbf{\Lambda}_{c_1}^{H} \mathbf{F}^{H} \mathbf{\Lambda}_{c_2}^{H} \mathbf{x},
\label{eqyhr02.5}
\end{equation}
where $\mathbf{F}$ is the discrete Fourier transform (DFT) matrix containing the element of $e^{j2\pi \frac{mn}{N}}$ at entry $(m,n)$, and $\mathbf{\Lambda_{c_i}}=\mathrm{diag}(e^{-j2{\pi}{c_i}n^2},n=0,1,...,N-1;i=1,2)$. To overcome the multipath effects and make the channel lies in a periodic domain, AFDM requires to add a chirp-periodic prefix (CPP), which is expressed as $s[n] = s[N + n] e^{-j2\pi c_1 (N^2 + 2Nn)}, \quad n = -N_{CPP}, \cdots, -1$, where $N_{CPP}$ is any integer greater than or equal to the maximum delay spread of the wireless channel.

For the doubly dispersive channel, the impulse response $g_n(\tau)$ is modeled as
\begin{equation}
g_n(\tau) = \sum_{i=1}^{P} h_i e^{-j\frac{2\pi}{N} \nu_i n} \delta\left(\tau - \tau_i\right),
\label{eqyhr02.6}
\end{equation}
where $P$ denotes the number of paths, $h_i,\tau_i, \nu_i$ denote the channel gain, delay shift and Doppler shift associated with $i_{th}$ path, respectively. 

After passing through the doubly dispersive channel and discarding the CPP, the received signals can be expressed in matrix form
\begin{equation}
\mathbf{r} = \mathbf{H} \mathbf{s} + \mathbf{w},
\label{eqyhr02.3}
\end{equation}
where $\mathbf{w} \sim \mathcal{CN}\left(0, N_0 \mathbf{I}\right)$ represents additive Gaussian noise and $\mathbf{H}$ is a $N\times N$ matrix denoted as 
\begin{equation}
\mathbf{H} = \sum_{i=1}^{P} h_i \mathbf{\Gamma}_{\rm CPP_i} \boldsymbol{\Delta}_{\nu_i} \mathbf{\Pi}^{\tau_i},
\label{eqyhr02.3}
\end{equation}
where $\mathbf{\Pi}^{\tau_i}$ is the forward cyclic-shift matrix, $\mathbf{\Delta}_{\nu_i} = \operatorname{diag}\left(e^{-j 2\pi \nu_i n}, n = 0, 1, \ldots, N - 1\right)$ is the Doppler shift matrix, and $\mathbf{\mathbf{\Gamma}_{\rm CPP_i}}$ is expressed as $
\Gamma_{\text{CPP}_i} = \text{diag}\left(
\begin{cases} 
e^{-j2\pi c_1 (N^2 - 2N(\tau_i - n))} & n < \tau_i \\
1 & n \geq \tau_i
\end{cases}
\right).
\label{eqyhr02.3}$

At the receiver side, the DAFT domain output symbols are obtained by
\begin{equation}
\begin{aligned}
\mathbf{y} &= \boldsymbol{\Lambda}_{c_2} \mathbf{F} \boldsymbol{\Lambda}_{c_1} \mathbf{r} \\
&= \sum_{i=1}^{P} h_i \boldsymbol{\Lambda}_{c_2} \mathbf{F} \boldsymbol{\Lambda}_{c_1} \boldsymbol{\Gamma}_{\rm CPP_i} \boldsymbol{\Delta}_{\nu_i} \boldsymbol{\Pi}^{\tau_i} \boldsymbol{\Lambda}_{c_1}^H \mathbf{F}^H \boldsymbol{\Lambda}_{c_2}^H \mathbf{x} + \widetilde{\mathbf{w}} \\
&= \mathbf{H}_{\rm eff} \mathbf{x} + \widetilde{\mathbf{w}},
\end{aligned}
\label{eqyhr02.3}
\end{equation}
where $\mathbf{H}_{\rm eff} \triangleq \boldsymbol{\Lambda}_{c_2} \mathbf{F} \boldsymbol{\Lambda}_{c_1} \mathbf{H} \boldsymbol{\Lambda}_{c_1}^H \mathbf{F}^H \boldsymbol{\Lambda}_{c_2}^H$ and $\widetilde{\mathbf{w}} = \boldsymbol{\Lambda}_{c_2} \mathbf{F} \boldsymbol{\Lambda}_{c_1} \mathbf{w}$.

While the AFDM waveform effectively captures full channel diversity in high-mobility environments, its performance remains critically depend on the accuracy of channel estimation\cite{b21,b22,b23,b24}. As the pilot can effectively facilitate channel estimation\cite{b25,b26}, in this work, we propose a novel Flag preamble structure and a Flag algorithm for low complexity and high accuracy channel estimation.

\section{TRADITIONAL AND PROPOSED FLAG BASED CHANNEL ESTIMATION}

\subsection{Traditional Flag-based Channel Estimation Algorithm}\label{AA}

This subsection details how the Flag algorithm exploits the unique peak-curtain property of the Flag preamble to decouple the conventional two-dimensional delay-Doppler search into two linear steps. 

We first introduce the unique peak-curtain property of the Flag preamble. After passing through the doubly dispersive channel, the channel introduces a shift to the Flag preamble, moving the correlation peak from the origin to $(\tau_0, \omega_0)$, representing the delay and Doppler shift. This results in AF known as the peak-curtain property

\begin{equation}
|\mathcal{M}(R, S_L)[\tau, \omega]| \approx \begin{cases} 1+\epsilon, & \text{if } (\tau, \omega) = (\tau_0, \omega_0) \quad  \\ 1/2+\epsilon, & \text{if } (\tau, \omega) \in L \setminus \{(\tau_0, \omega_0) \} \quad  \\ \epsilon, & \text{otherwise} \end{cases}
\label{eqyhr02.3}
\end{equation}
where $S_L$ and $R$ denote the transmitter and the receiver sequences, $\mathcal{M}(R, S_L)$ denotes the AF of $S_L$ and $R$, $\epsilon < 1/\sqrt{N}$, symbol `` $\setminus$ " denotes  exception.

Due to the peak-curtain property mentioned above, channel estimation is equivalent to a peak-finding problem, which is solved in two steps\cite{b13}:

Step 1 (Line Search): Select a line $L_{\perp}$ transversal to the original Flag preamble $L$. Compute $\mathcal{M}(R, S_L)[\tau, \omega]$ with $(\tau, \omega) \in L_{\perp}$ by fast Fourier transform (FFT), constituting a linear search. Identify $(\tau', \omega')$ such that $||\mathcal{M}(R, S_L)[\tau', \omega']|-1/2|<\epsilon$, finding curtain at the shifted line $L_{\perp}$ of $L$.

Step 2 (Peak Search): Compute $\mathcal{M}(R, S_L)$ on $L+(\tau_0, \omega_0)$ and find $(\tau'', \omega'')$ such that $||\mathcal{M}(R, S_L)[\tau'', \omega'']|-1|<\epsilon$, i.e., finding the peak at $(\tau_0, \omega_0)$, that is the estimated delay and Doppler.

However, the traditional Flag algorithm remains susceptible to false alarms and missed detections, particularly when 

confronted with extended targets, environmental clutter and occasional obscuration of weak targets by the curtains of strong targets\cite{b13}. To mitigate the probabilities of false alarms and missed detections, we propose an enhanced Flag algorithm for channel estimation in the next subsection.

\subsection{Proposed Flag-based Channel Estimation Algorithm}
In the conventional Flag algorithm based on HWS \cite{b11}, the ideal sidelobe properties of prime-length sequences allow for a straightforward peak detection. However, for arbitrary-length sequences where $N$ is non-prime, the AF of the optimized Flag preamble may exhibit non-ideal sidelobes due to the non-ideal Flag sequence. 
These sidelobes can be erroneously identified as true paths, leading to false peaks and severe performance degradation. 
To address this, we propose an enhanced estimation algorithm incorporating a candidate selection strategy and global LS refinement. The general framework of the proposed Flag algorithm is outlined in \textbf{Algorithm 1}, and the detailed implementation steps are elaborated as follows.

{
\setlength{\textfloatsep}{-0.7cm}
\begin{algorithm}[t]
\caption{Proposed Flag-based Channel Estimation Algorithm}
\label{alg:robust_estimation}
\begin{algorithmic}[1]
\REQUIRE Received Flag preamble $\mathbf{r}$, Transmitted Flag preamble $\mathbf{f}$, Path number $P$, Candidates $K$, Threshold $\gamma$.
\ENSURE Channel parameters $\{(\hat{\tau}_i, \hat{\nu}_i, \hat{h}_i)\}_{i=1}^P$.
\STATE \textbf{Phase 1: Candidate-Aided Line Search}
\FOR{$i = 1$ \TO $P$}
    \STATE $\mathbf{a} = \text{FFT}(\mathbf{r} \odot \mathbf{f}^* e^{-j\frac{2\pi kn}{N} })$ \COMMENT{\textit{Line Search}}
    \STATE $\hat{\mathcal{K}} \leftarrow \{ \text{indices of top } K \text{ peaks } > \gamma \cdot \max |\mathbf{a}| \}$
    \FOR{\textbf{each} $K \in \hat{\mathcal{K}}$}
        \STATE $\mathbf{z}_k = \text{IFFT} \left\{ \text{FFT}(\mathbf{t}_k[n] \odot \mathbf{d}[n]) \odot \text{FFT}(\mathbf{f}[n] \odot \mathbf{d}[n])^* \right\}$ \COMMENT{\textit{Peak Search}}
    \ENDFOR
    \STATE Select $(\hat{\tau}_i, \hat{\nu}_i) = \arg \max_{k, \tau} |\mathbf{z}_k|$
\ENDFOR
\STATE \textbf{Phase 2: Global Refinement}
\STATE Construct basis matrix $\boldsymbol{\Phi} \in \mathbb{C}^{N \times P}$ using all detected $(\hat{\tau}_i, \hat{\nu}_i)$.
\STATE $\hat{\boldsymbol{h}} = (\boldsymbol{\Phi}^H \boldsymbol{\Phi})^{-1} \boldsymbol{\Phi}^H \mathbf{r}$ \COMMENT{\textit{Least Squares Estimation}}
\RETURN $\{(\hat{\tau}_i, \hat{\nu}_i, \hat{h}_i)\}_{i=1}^P$
\end{algorithmic}
\end{algorithm}
}

\subsubsection{Multi-Candidate Line Search}
The proposed algorithm decouples the 2D delay-Doppler search into a near-linear search. 
In the line search stage, the received preamble signal $\mathbf{r}$ is first cross-correlated with the locally generated Flag preamble $\mathbf{f}$, followed by FFT as
\begin{equation}
\mathbf{a}[k] = \text{FFT}\left\{\left| \sum_{n=0}^{N-1} \mathbf{r}[n] \mathbf{f}^*[n] e^{-j\frac{2\pi kn}{N}} \right|\right\}, \quad k = 0, \dots, N-1,
\end{equation}
where $N$ is the length of the preamble.

To prevent the algorithm from prematurely locking onto a strong sidelobe which maybe a pseudo-peak that causes the missed detections, we define a candidate set $\hat{\mathcal{K}}$ containing $K$ potential indices
\begin{equation}
\hat{\mathcal{K}}= \{ k \mid \mathbf{a}[k] \geq \gamma \cdot \max_{k} \mathbf{a}[k], \text{ and } \mathbf{a}[k] \in \text{Top-}K \text{ values} \},
\end{equation}
where $\gamma$ is the relative energy threshold and $K$ is the candidate number. 
This strategy ensures that even if a true path's intercept is partially masked by noise or sidelobes, it can effectively prevent the missed detections of true peaks.

\subsubsection{Peak Search}
Each candidate $K \in \hat{\mathcal{K}}$ obtained from the line search stage is subjected to the peak search stage. The formula of this stage is 
\begin{equation}
\mathbf{z}_k = \text{IFFT} \left\{ \text{FFT}(\mathbf{t}_k[n] \odot \mathbf{d}[n]) \odot \text{FFT}(\mathbf{f}[n] \odot \mathbf{d}[n])^* \right\},
\end{equation}
where $\mathbf{t}_k[n] = \mathbf{r}[n] e^{-j \frac{2\pi \nu_k n}{ N}}$ is the Doppler-compensated signal of the received signal $\mathbf{r}[n]$ at the intercept $k$, 
and $\mathbf{d}[n] = e^{-j\frac{\pi \xi n^2}{N}}$ is the de-chirp vector, $\mathbf{z}_k$ is the AF of the received Flag preamble evaluated along the shifted line $L + (\tau_0, \omega_0)$, and $\xi$ is the slope of the line $L$ in the time-frequency plane $V$, it represents the rate of frequency change over time for the Heisenberg sequence of the Flag preamble.
The candidate yielding the maximum correlation peak across all $K$ hypotheses is recorded as the $i$-th path $(\hat{\tau}_i, \hat{\nu}_i)$.

\subsubsection{Simulation-based Parameter Optimization}
The selection of $\gamma$ and $K$ involves a fundamental trade-off among the probability of detection (PD), probability of false alarm (PFA), and probability of miss (PM). 
In our simulation framework, we perform a comprehensive parameter sweep under various signal-to-noise ratio (SNR) conditions. 
\begin{itemize}
    \item Candidate number $K$: Simulations indicate that the PD of the strongest path reaches a stable plateau when $K \geq 3$. Further increasing $K$ will unnecessarily increase the complexity of Phase 2 without significantly performance improvement.
    \item Relative threshold $\gamma$: By analyzing PM and PFA trade-off, a threshold of $\gamma \approx 0.2$ to $0.3$ is found to be optimal. This range effectively filters out ambient noise and side-peaks while preserving potential paths in high-mobility channels.
\end{itemize}

\subsubsection{Global LS Refinement}

After identifying the parameters of all $P$ paths, the initial complex gains obtained via successive interference cancellation (SIC) may still suffer from residual errors. 

To deal with this problem, we perform a joint refinement by constructing a basis matrix $\boldsymbol{\Phi} \in \mathbb{C}^{N \times P}$, where $N$ denotes the length of the Flag preamble and each column $\boldsymbol{\phi}_i$ represents the reconstructed time-domain signal for the $i$-th path, defined as
\begin{equation}
\boldsymbol{\phi}_i = \mathbf{S}(\hat{\tau}_i, \hat{\nu}_i) \mathbf{f} = \mathbf{f}[n-\hat{\tau}_i] e^{j\frac{2\pi \hat{\nu}_i n}{N}},
\end{equation}
where $\mathbf{f} \in \mathbb{C}^{N \times 1}$ denotes the original Flag preamble of length $N$, $\hat{\tau}_i$ is the estimated time delay index for the $i$-th path, $\hat{\nu}_i$ is the estimated normalized Doppler shift index for the $i$-th path, $\mathbf{S}(\hat{\tau}_i, \hat{\nu}_i)$ is the delay-Doppler shift operator, mathematically equivalent to the Heisenberg operator $\pi(\tau, \omega)$\cite{b10}, which performs a cyclic shift and phase rotation on the sequence. The complex gain vector $\hat{\boldsymbol{h}} = [\hat{h}_1, \dots, \hat{h}_P]^T$ is then determined by solving the following optimization problem
\begin{equation}
\min_{\hat{\boldsymbol{h}}} \| \mathbf{r} - \boldsymbol{\Phi}\hat{\boldsymbol{h}} \|^2,
\end{equation}
where $\mathbf{r} \in \mathbb{C}^{N \times 1}$ denotes the received discrete-time preamble vector.

The closed-form solution is given by the Moore-Penrose pseudoinverse
\begin{equation}
\hat{\boldsymbol{h}} = (\boldsymbol{\Phi}^H \boldsymbol{\Phi})^{-1} \boldsymbol{\Phi}^H \mathbf{r}.
\end{equation}
This step ensures that the gains of all paths are obtained simultaneously, which is particularly critical for non-prime length preambles to suppress inter-path leakage and achieve a low bit error rate (BER).

\subsection{Complexity Analysis}
Traditional doubly selective channel estimation via maximum likelihood (ML) or MF typically incurs a prohibitive complexity of $\mathcal{O}(N^2 \log N)$ \cite{b13}. 
In contrast, our proposed algorithm achieves a near-linear complexity of $\mathcal{O}(KPN \log N)$. Specifically, for each path, Phase 1 identifies the candidate set $\hat{\mathcal{K}}$ with $\mathcal{O}(N \log N)$ operations, while Phase 2 performs verification for $K$ hypotheses with $\mathcal{O}(KN \log N)$ cost. Since $K$ and the number of paths $P$ are small constants independent of $N$, the overall complexity is dominated by $\mathcal{O}(N \log N)$. Furthermore, our optimization-based design maintains this high efficiency for arbitrary $N$, achieving a superior balance between accuracy and real-time latency.

\section{SIMULATION RESULTS AND DISCUSSION}

In this section, we evaluate the channel estimation performance of the traditional Flag preamble and the proposed Flag preamble algorithms. 
We first analyze the parameter selection for the candidate selection strategy. 
Subsequently, by taking AFDM as an example, we evaluate the traditional Flag algorithm and the proposed Flag algorithm on their mean square error (MSE) performance, and give the PD and PM to show the superiority of the proposed Flag algorithm. 
Finally, for the AFDM waveform, we test the BER performance. 
These comparisons demonstrate that the proposed Flag algorithm achieves superior channel estimation accuracy compared to the traditional algorithm, while effectively mitigating the complex sidelobe interference inherent.

\subsection{Simulation Parameters}
To evaluate the performance of the proposed Flag-based channel estimation, the simulation considers a high-mobility scenario with a carrier frequency of $f_c = 4$~GHz. To simulate extreme Doppler effects, the maximum user equipment (UE) speed is set to $540$~kmph, resulting in a maximum Doppler shift of $2$~kHz.  Specifically, we employ a simplified version of the extended vehicular A channel model, which is configured with 4 independent taps.The key simulation parameters are summarized in TABLE~\ref{tab:sim_parameters}.

\begin{table}[t]
    \caption{System Simulation Parameters}
    \label{tab:sim_parameters}
    \centering
    \begin{tabular}{|l|l|}
        \hline
        \textbf{Parameter} & \textbf{Value} \\ \hline
        Carrier frequency $f_c$ & 4 GHz \\ \hline
        AFDM subcarrier spacing $\Delta f_{\text{AFDM}}$ & 1 kHz \\ \hline
        AFDM subcarrier number (frame size) $N_{\text{AFDM}}$ & 1024 \\ \hline
        Bandwidth $B_{\text{AFDM}}$ & 1024 kHz \\ \hline
        Maximum UE speed & 540 kmph \\ \hline
        Maximum Doppler shift & 2 kHz \\ \hline
        Number of paths ($P$) & 4 \\ \hline
        Modulation scheme & 4-QAM \\ \hline
        SISO configuration & $1 \times 1$ \\ \hline
        Detector type & LMMSE \\ \hline
    \end{tabular}
    \vspace{- 0.3 cm}
\end{table}

\subsection{Parameters Selection for the Proposed Flag Algorithm}
The effectiveness of the Flag algorithm in non-prime scenarios depends heavily on the candidate set $\hat{\mathcal{K}}$. 
To establish an optimal baseline, we analyze the impact of the candidate number $K$ and the relative threshold $\gamma$ through a comprehensive parameter sweep.

\begin{figure}[t]
    \centering
    \begin{subfigure}[b]{0.19\textwidth}
        \centering
        \includegraphics[width=\textwidth]{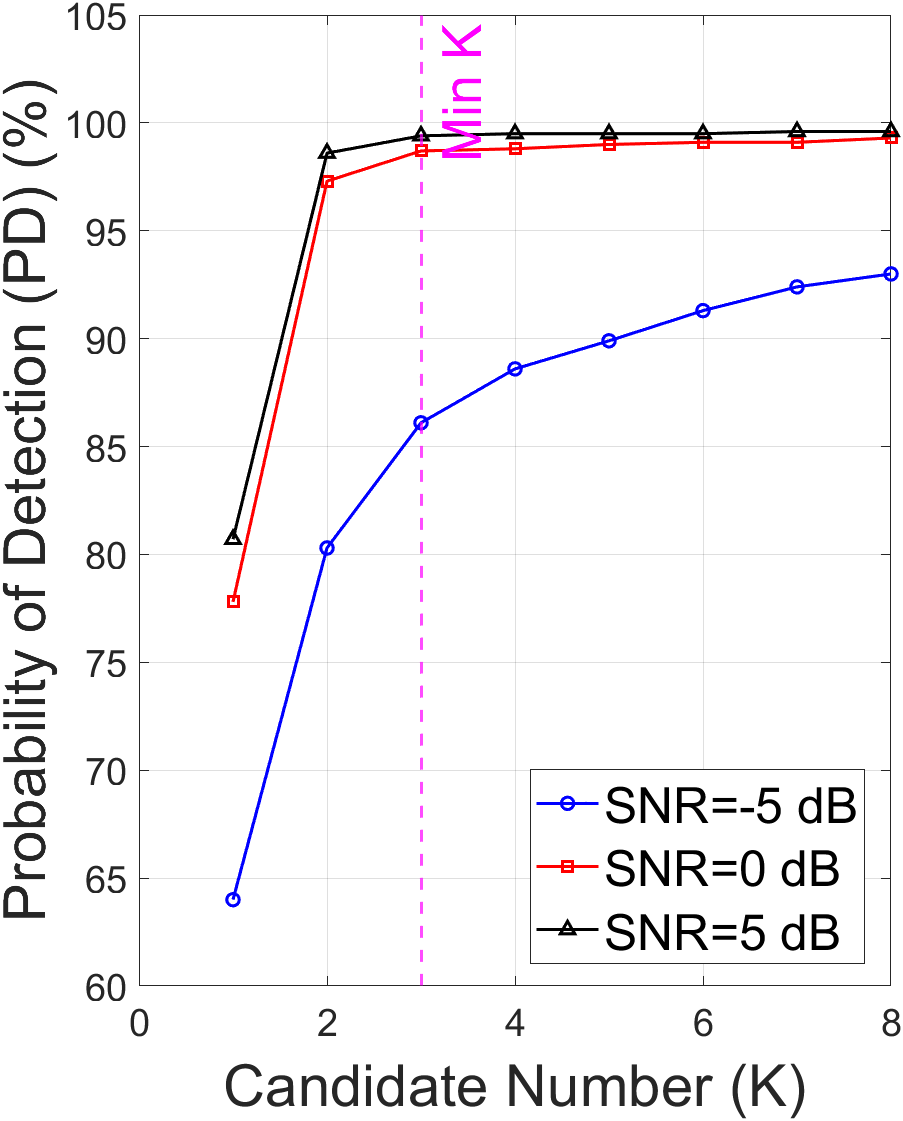}
        \caption{Impact of $K$}
        \label{fig:parameter_selection_M}
    \end{subfigure}
    \hfill 
    \begin{subfigure}[b]{0.223\textwidth}
        \centering
        \includegraphics[width=\textwidth]{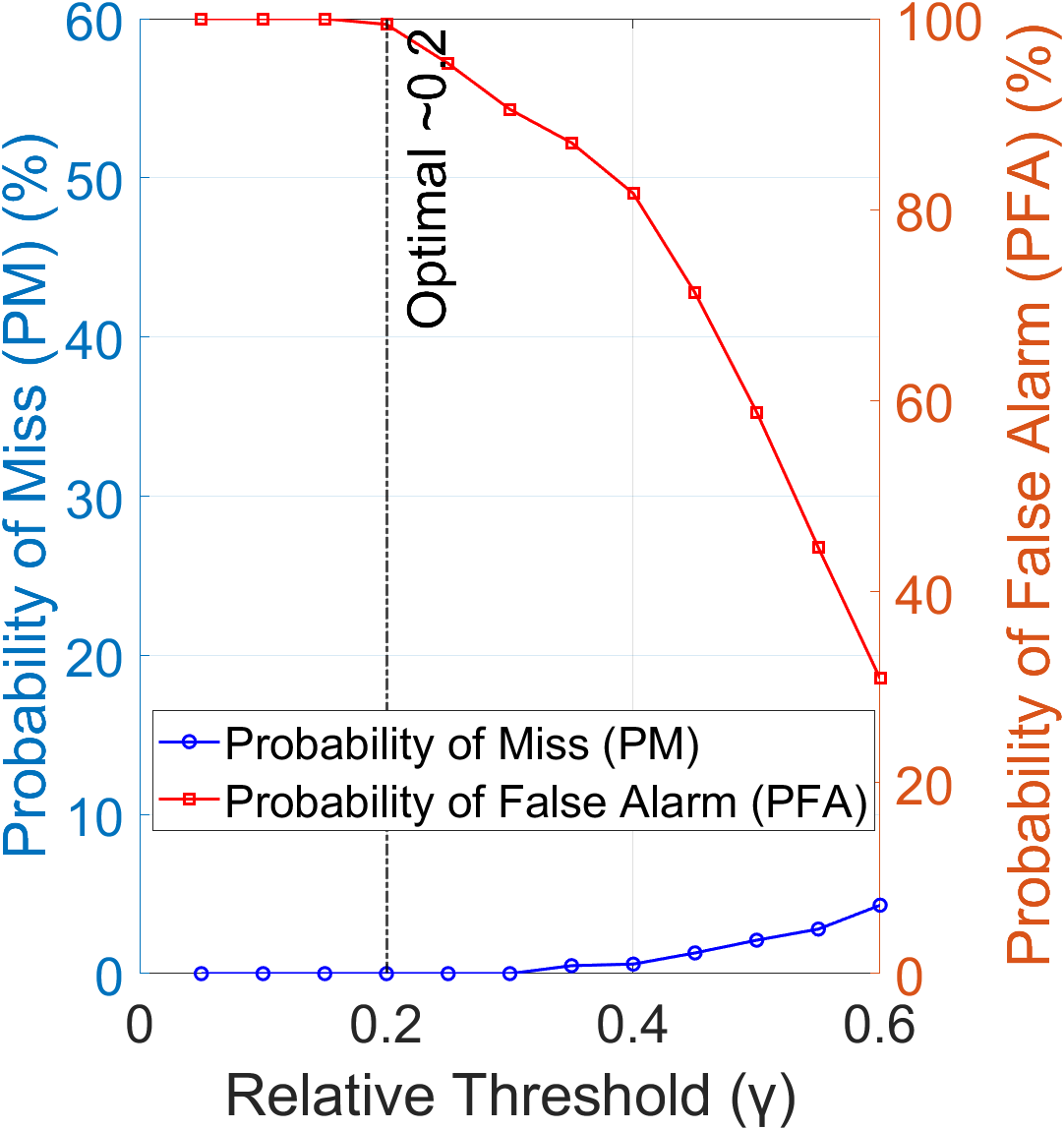}
        \caption{Impact of $\gamma$}
        \label{fig:parameter_selection_gamma}
    \end{subfigure}
    
    \caption{Parameters selection for the proposed Flag algorithm: (a) Trade-off regarding candidate numbers $K$; (b) Trade-off regarding relative threshold $\gamma$.}
    \label{fig:overall_parameter_selection}
    \vspace{- 0.6 cm}
\end{figure}

\begin{itemize}
    \item Detection trade-offs: Fig. \ref{fig:parameter_selection_M} depicts the PD of the strongest path across various SNR levels. In Fig. \ref{fig:parameter_selection_M}, we can see that the PD scales positively with the SNR and eventually levels off in the high-SNR regime. The results also confirm that the probability of the true intercept being included in the candidate set reaches a stable plateau when $K \geq 3$. This justifies our choice of a small candidate set, which preserves the $\mathcal{O}(N \log N)$ complexity of the algorithm. 
     \item Candidate efficiency: Fig. \ref{fig:parameter_selection_gamma}  illustrates the trade-off between the PM and PFA as a function of the relative threshold $\gamma$. It is observed that a threshold range of $\gamma \in [0.2, 0.3]$ effectively filters out false alarms from false peaks while maintaining high sensitivity for true paths.
    
\end{itemize}

In summary, the parameter sweep identifies an optimal configuration of $K \geq 3$ and $\gamma \in [0.2, 0.3]$. This setup strikes a critical balance between detection reliability and computational efficiency, providing a robust foundation for evaluating the system's performance.

\subsection{Performance of MSE on AFDM}

In this subsection, we investigate the performance of the proposed Flag algorithm with candidate selection strategy. For AFDM, we use the traditional Flag algorithm and the proposed Flag algorithm to establish the estimated channel matrix, calculating the MSE between the real channel matrix and the estimated channel matrix, respectively. We then analyze the PD and PM to maintain high accuracy and low resource consumption.

\begin{figure}[t]
        \centering
        \includegraphics[width=0.35\textwidth]{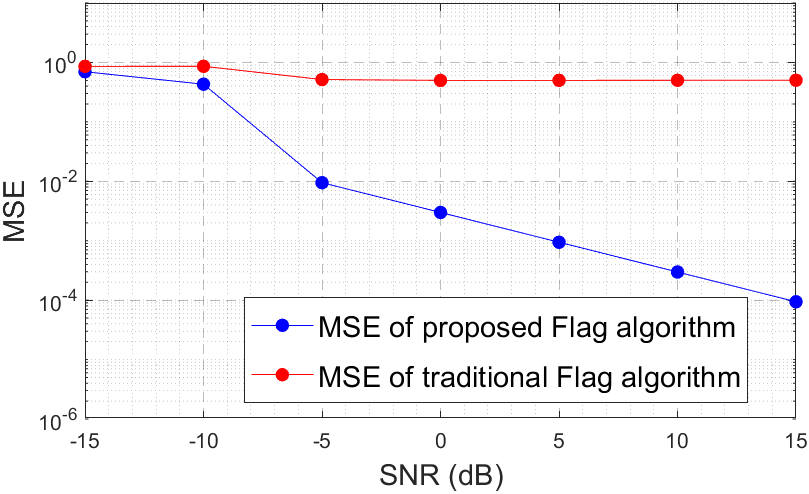}
        \caption{MSE results for channel estimation algorithms.}
        \label{fig:MSE_final}
    \vspace{ -0.5 cm}
\end{figure}

\begin{figure}[t]
        \centering
        \includegraphics[width=0.35\textwidth]{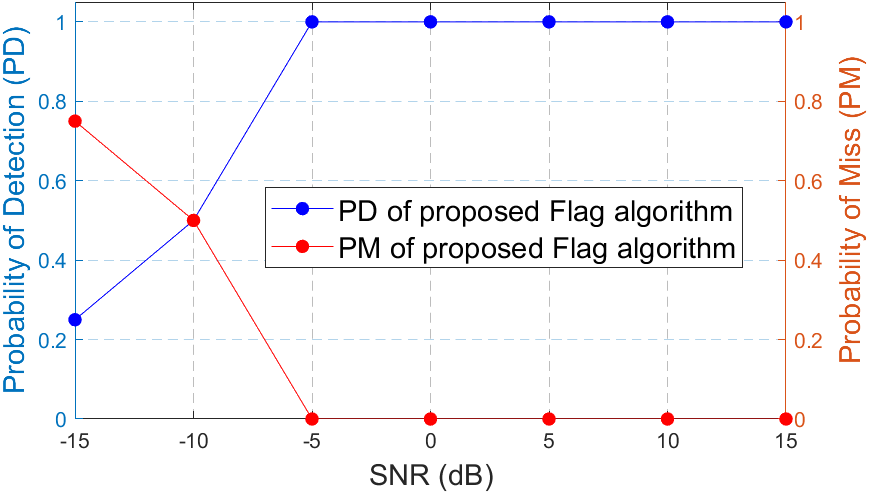}
        \caption{PD and PM of proposed Flag algorithm.}
        \label{fig:PD_final}
        \vspace{ -0.6 cm}
\end{figure}

\begin{itemize}
\item MSE performance of proposed Flag algorithm on AFDM: In AFDM framework, we evaluate the impact of the proposed Flag algorithm and the traditional Flag algorithm on the MSE performance. As shown in Fig. \ref{fig:MSE_final}, the red line represents the MSE of the estimated channel matrix calculated by the traditional Flag algorithm compared to the true channel matrix, while the blue line represents the MSE of the estimated channel matrix calculated by the proposed Flag algorithm. The MSE for each Monte Carlo trial is defined as
\begin{equation}
    \text{MSE} = \frac{\|\mathbf{H} - \mathbf{\hat{H}}\|_F^2}{\|\mathbf{H}\|_F^2},
\end{equation}
where $\mathbf{H}$ and $\mathbf{\hat{H}}$ denote the actual and estimated 
delay-Doppler domain effective channel matrices, respectively, and 
$\|\cdot\|_F$ denotes the Frobenius norm. This metric encapsulates the 
compounded errors arising from path parameter acquisition and complex 
gain recovery. In our simulations, we can see that the proposed Flag algorithm achieves greater accuracy than the traditional one. 

\item The PD of Proposed Flag Algorithm: In Fig. \ref{fig:PD_final}, note that as the SNR increases, the PD also increases and then tends to saturate; in contrast, the PM decreases and then tends to saturate. Considering both PD and PM, we can see that with the SNR increasing, the proposed Flag algorithm can get full detection, which shows the algorithm has promising performance.
\end{itemize}

In this subsection, the proposed Flag algorithm consistently outperforms the traditional version in channel estimation accuracy. By analyzing the trade-off between PD and PM, it is observed that the proposed Flag algorithm provides an optimal balance between resource efficiency and channel acquisition precision.

\subsection{Performance of BER on AFDM}
Subsequently, taking AFDM  \cite{b18,b19,b27,b28,b29} as an exmaple, we investigate the BER performance by using our proposed channel estimation algorithm, traditional channel estimation algorithm and compare with the perfect CSI, to demonstrate the universality of the proposed channel estimation algorithm.

\begin{figure}[t]

        \centering
        \includegraphics[width=0.371\textwidth]{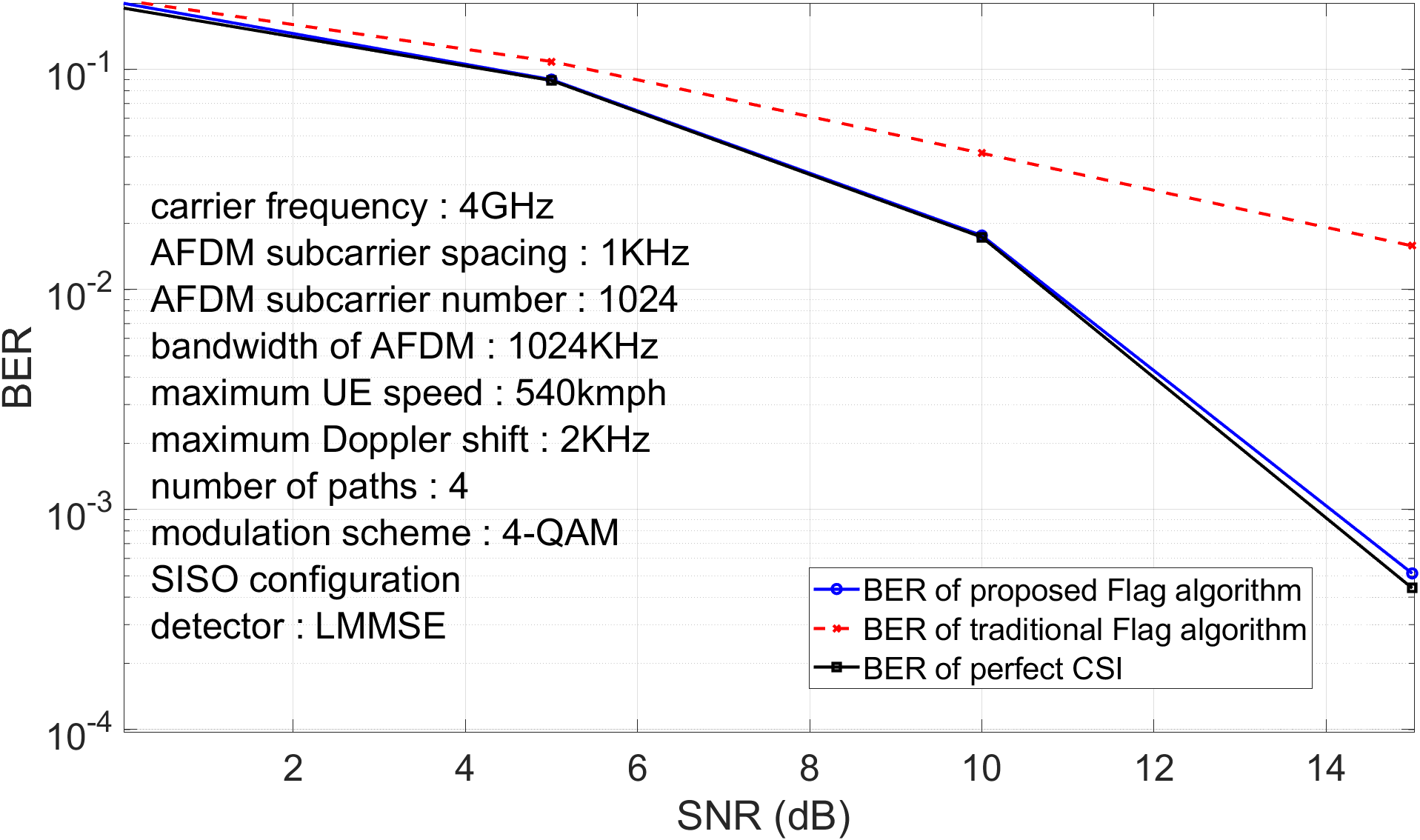}
        \caption{BER results for channel estimation algorithms.}
        \label{fig:BER_FINAL}
        \vspace{ -0.8 cm}
\end{figure}

\begin{itemize}
\item BER performance of proposed Flag algorithm on AFDM: To verify the universality of the proposed scheme, we integrate the Flag preamble into AFDM. The simulation is conducted under a highly dynamic extended vehicular A(EVA) channel with a maximum Doppler shift of 2 kHz, corresponding to a UE speed of $540 $ kmph at $f_c$ = 4 GHz.
As illustrated in Fig. \ref{fig:BER_FINAL}, the simulation results demonstrate that the BER performance of the proposed Flag algorithm is remarkably close to the perfect CSI baseline across the entire SNR range, and shows greater BER performance than the traditional Flag algorithm. As AFDM is a unified framework that can compatible waveforms like OFDM, OCDM, etc, we can extend our proposed Flag preamble channel estimation algorithm into other NEWs in a straightforward manner.
\end{itemize}

Overall, we can see from the results that the proposed Flag algorithm offers significant advantages for channel estimation due to its near-linear complexity and high accuracy.

\vspace{ -0.13 cm}
\section{CONCLUSION}
In this paper, we have proposed a Flag algorithm with a candidate selection strategy that enables robust delay-Doppler estimation for arbitrary sequence lengths, and we applied this proposed Flag preamble algorithm for channel estimation in AFDM systems. In the simulation results, the proposed Flag algorithm achieves near-optimal BER performance with nearly linear complexity, which provides a new preamble scheme for channel estimation with low complexity and high accuracy.

\vspace{ -0.13 cm}

\end{document}